\documentclass[preprint,preprintnumbers,amsmath,amssymb]{revtex4}

\usepackage{graphicx}
\usepackage{dcolumn}
\usepackage{bm}

\begin{document}

\title{Environmental Metal Pollution Considered as Noise: Effects on the Spatial Distribution of Benthic Foraminifera in two Coastal Marine Areas of Sicily (Southern Italy)}
\author{D. Valenti\footnote{Corresponding author: valentid@gip.dft.unipa.it}, L. Tranchina, M. Brai, A. Caruso, C. Cosentino, B. Spagnolo}
\affiliation{Dipartimento di Fisica e Tecnologie Relative,
Universit\`a di Palermo and Unità CNISM Palermo \\Viale delle
Scienze, Edificio 18, I-90128 Palermo, Italy}

\begin{abstract}
\noindent We analyze the spatial distributions of two groups of
benthic foraminifera (\emph{Adelosina} spp. + \emph{Quinqueloculina}
spp. and \emph{Elphidium} spp.), along Sicilian coast, and their
correlation with six different heavy metals, responsible for the
pollution. Samples were collected inside the Gulf of Palermo, which
has a high level of pollution due to heavy metals, and along the
coast of Lampedusa island (Sicily Channel, Southern Mediterranean),
which is characterized by unpolluted sea waters. Because of the
environmental pollution we find: (i) an anticorrelated spatial
behaviour between the two groups of benthic foraminifera analyzed;
(ii) an anticorrelated (correlated) spatial behaviour between the
first (second) group of benthic foraminifera with metal
concentrations; (iii) an almost uncorrelated spatial behaviour
between low concentrations of metals and the first group of
foraminifera in clean sea water sites. We introduce a two-species
model based on the generalized Lotka-Volterra equations in the
presence of a multiplicative noise, which models the interaction
between species and environmental pollution due to the presence in
top-soft sediments of heavy metals. The interaction coefficients
between the two species are kept constant with values in the
coexistence regime. Using proper values for the initial conditions
and the model parameters, we find for the two species a theoretical
spatial distribution behaviour in a good agreement with the data
obtained from the $63$ sites analyzed in our study.\\ \\ Keywords:
Heavy metals, Benthic foraminifera, Lotka–Volterra, Population
dynamics, Multiplicative noise, Noise-induced phenomena
\end{abstract}

\maketitle

\section{Introduction}\label{Introduction}

Natural ecosystems are open complex systems in which the intrinsic
nonlinearity together with the environmental interaction gives rise
to a rich variety of experimental phenomenologies. These can be
explained through some recent counterintuitive noise-induced
phenomena like stochastic resonance, noise enhanced stability and
noise delayed extinction (Spagnolo et al., 2004). Random
fluctuations and their role in natural systems have become an
important topic of investigation in many different fields ranging
from neuroscience to biological evolution and population dynamics
(Goldenfeld and Kadanoff, 1999; Freund and P$\ddot{o}$schel, 2000;
Brown et al., 2001, Sugden and Stone, 2001; Turchin et al., 2002;
Spagnolo et al., 2003; Spagnolo et al., 2004; Valenti et al.,
2004a). Moreover noise-induced effects in biological systems appear
in bio-informatics (Blake et al., 2003; Ozbudak et al., 2002), virus
dynamics, epidemics (Tuckwell and Le Corfec, 1998; Gielen, 2000;
Chichigina et al., 2005) and population dynamics for zooplankton
abundances (Caruso et al., 2005). Specifically the role of the noise
in population dynamics has been intensively analyzed in theoretical
investigations (Ciuchi et al., 1996; Vilar and Sol\'e, 1998;
Giardina et al., 2001; Rozenfeld et al., 2001; Scheffer et al.,
2001; Staliunas, 2001; La Barbera and Spagnolo, 2002; Spagnolo and
La Barbera, 2002a; Spagnolo et al., 2002b; Cirone et al., 2003).\\
Experimental data for population dynamics are often related to the
spatial extension of the system considered. Understanding the
dynamics which determines the spatial structures, that is the
spatio-temporal patterns, is an important target in the analysis of
the marine ecological time series in view of modelling their complex
behaviour (Lourens et al., 1996; Sprovieri et al., 2003; Caruso,
2004 and references therein). In this context it is fundamental to
understand the effects of noise in order to describe the
spatio-temporal dynamics of real ecosystems (Zhonghuai et al., 1998;
Blasius et al., 1999; King et al., 2001; Rozenfeld et al., 2001;
Valenti et al., 2004b; Valenti et al., 2006).\\
In this paper we analyze abundances of two groups of benthic
foraminifera and concentrations of six heavy metals. Data were
obtained from $58$ sites inside the Gulf of Palermo, containing high
levels of pollutants (heavy metals), and from $5$ sites along the
coast of the Lampedusa island (Sicily Channel) characterized by
unpolluted or very low polluted sea waters. In order to check
whether the presence of heavy metals in the top-soft sediments
affects the spatio-temporal dynamics of the two groups of benthic
foraminifera, data were analyzed calculating spatial correlations
among species abundances and metal concentrations. Finally, to
understand the spatial distributions of the two groups of
foraminifera interacting with each other through the presence of the
heavy metals in the environment, we consider a model of two
generalized Lotka-Volterra equations in the presence of two peculiar
multiplicative noises (Valenti et al., 2004a). The noise sources
take into account for the interaction between the species and the
heavy metal pollution. In the final section of the paper, the
theoretical spatial behaviour obtained with this model are compared
with the experimental ones obtained from samples of the considered
sites.

\section{Materials and methods}

\subsection{Sampling methods}

We carried out the study on $63$ samples taken from top
soft-sediments, collected between autumn $2004$ and summer $2006$
using a Van Veen grab. Sampling site locations have been determined
with Garmin $12$ channels GPS (Global Position System). In the Gulf
of Palermo (Sicily, Italy) a total of $58$ samples sub-divided in
$24$ stations, far from the coast line, have been considered.
Generally each station consists of three samples taken at different
bathymetries. The $58$ samples collected in the Gulf of Palermo are
shown in Fig.1. It is possible to distinguish three different
"strips" formed by the sites (black dots) where the samples were
collected. From the coast towards the shelf area one can recognize
an \emph{internal} strip (GP X-1), an \emph{intermediate} one (GP
X-2), and an \emph{external} one (GP X-3), whose bathymetries are
respectively $-10~m$, $-20~m$, $-30~m$. The sites of the Gulf of
Palermo were named "GP" followed by two numbers: the first one for
the position inside the strip, the second one for the numeration of
the strip ($1$, $2$ or $3$). For example, "GP~$9$-$1$" means "site
$9$ belonging to strip $1$ in Gulf of Palermo". Along the coast of
Lampedusa Island we collected randomly $5$ samples with different
bathymetries ranging from $-3~m$ to $-30~m$.

\subsection{Benthic foraminifera}

For the analytical studies only the uppermost part of soft-bottom
sediments (3-4 cm) was utilized. One hundred grams of wet sediment
of each sample were oven-dried at 80 $^\circ$C for 48 hours; the
dried sediment was weighted and washed through a 63 $\mu$m sieve.
Quantitative analysis on benthic foraminifera was carried out on
dead assemblage fraction $>$ 63 $\mu$m, focusing on thanatocoenosis,
using an Otto microsplitter to obtain a smaller, but statistically
representative, amount of the total sediment. The total number of
benthic foraminifera, contained in this fraction, has been counted.
The accuracy in the counting of each species is less than 10\%. The
taxonomy of benthic foraminifera was essentially based on the
classifications by Loeblich and Tappan (1964, 1988), and Sgarrella
and Montcharmont Zei (1993). For this study we focused our attention
on the percentages of some genera of miliolids (\emph{Adelosina}
spp. and \emph{Quinqueloculina} spp.) and the genus
\emph{Elphidium}. Typically these genera are abundant in the top
sediments of the upper shelf zone, but in the marine ecosystem the
dominance of one of this group depends on chemical parameters as
oxygen, salinity and nutrient concentrations. In particular
miliolids generally prefer high oxygen concentration in the shelf
area waters while \emph{Elphidium} spp. are more tolerant to
stressed environmental conditions for changes in salinity  and high
levels of nutrients (Sen Gupta, 2003).

\subsection{Metal measurements}

The heavy metals considered are zinc ($Zn$), copper ($Cu$), chrome
($Cr$), iron ($Fe$), lead ($Pb$), mercury ($Hg$). For Flame Atomic
Absorption Spectrophotometry "pseudo-total metal contents" analysis,
a quantity of 1000 mg was digested in an open cavity microwave
system (C.E.M. Star system 2) using the following procedure: 20 mL
of HNO$_3$ 65\% and 10 mL of H$_2$SO$_4$ 96\% were first added to
the 1000 mg sample aliquot and heated for five minutes at 75
$^\circ$C. Temperature was then raised up to 85 $^\circ$C and kept
for ten minutes. Afterward temperature was raised to 95° C for ten
minutes, then kept at 106 $^\circ$C for seven minutes, at 115
$^\circ$C for fifteen minutes, at 120 $^\circ$C for ten minutes and
at the boiling point for fifteen minutes (Man et al., 2004).
Afterward, 15 mL of H$_2$O$_2$ 30\% were added and the last above
described four temperature steps were repeated. All reagent were of
Merck Suprapure grade. The term "pseudo-total" (Manta et al., 2002)
accounts for the used digestion procedure not completely destroying
silicates (Cook et al., 1997). Digested samples were cooled,
filtered through 0.45 $\mu$m pores, and diluted to 100 mL with water
(resistivity 18 M  cm Smeg $WP4100/A10$). Cr, Cu, Fe, Pb and Zn
concentrations were measured by a Varian SpectraAA $220$ FS. The
spectrophotometer was equipped with a deuterium background
corrector. Cr, Cu and Pb were measured using the standard addition
methods, Zn was measured after that a calibration curve was
performed and diluting samples 1 to 11. Hg was measured using the
Varian SpectrAA $220$ FS coupled with the continuous flow vapor
generator (Varian VGA-77) and a SnCl$_2$ solution as reducing agent
for Hg vapor release from sample solutions. Working standard
solutions of metals were prepared using standard solution of each
metal $1000$ mg L$^{-1}$ (Merck). All samples were analyzed in
duplicate. All glassware were previously soaked overnight with 10\%
HNO$_3$ solution and then rinsed with distilled and deionized water.
The National Research Council Canada PACS-2 (Marine sediment) was
used as certified reference material to test the repeatability of
the measurements and to evaluate recovery (Table 1). For some metals
(Cr and Pb) recovery was not complete because digestion methods used
do not destroy silicates. The measurements are expressed in weight
of metal per weight of dry sediment. The accuracy is about $7\%$.
\begin{table}[htbp]
\begin{center}
\includegraphics[width=14cm]{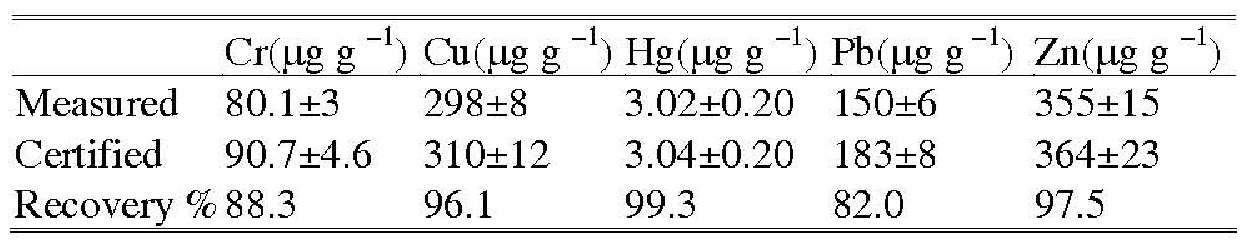}
\end{center}
\caption{ \small \emph{Results expressed as mean $\pm$ standard
deviation of 10 measures obtained by measurements of certified
reference material.}\medskip}
 \label{table_correlation_LAMP_2}
\end{table}

\section{Experimental data}\label{Experimental_data}

\subsection{Data from Gulf of Palermo}\label{Palermo_data}
\begin{figure}[htbp]
\begin{center}
\includegraphics[width=12cm]{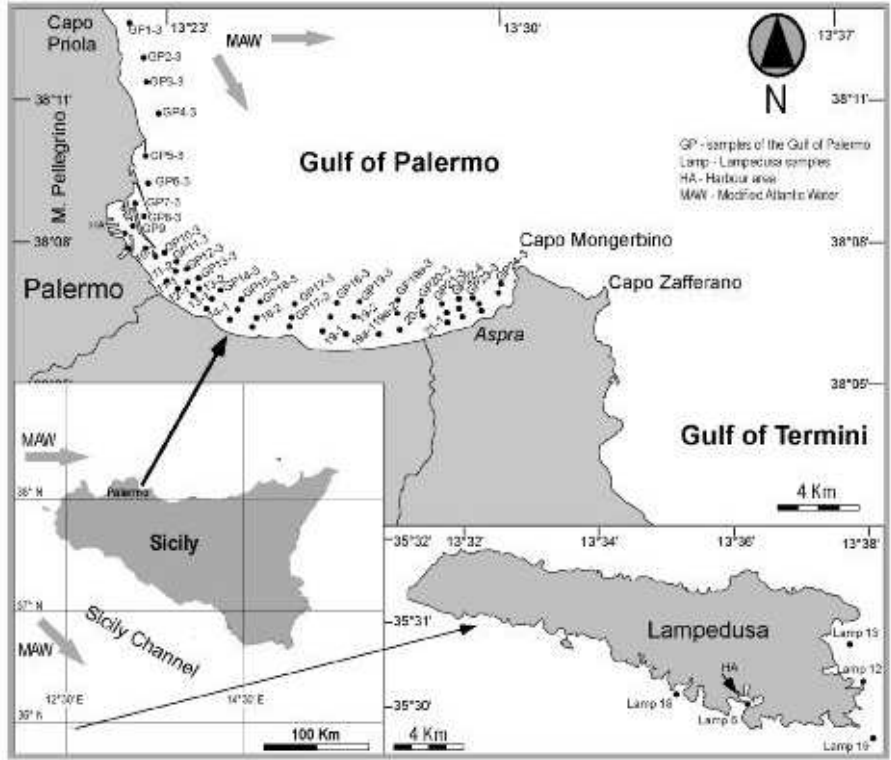}
\end{center}
\caption{ \small \emph{Gulf of Palermo. The black dots indicates the
sites where the samples were collected. The sites "draw" three
different "strips" formed by the sites (black dots) where the
samples were taken. From the coast towards the shelf area one can
recognize an \emph{internal} strip (GP X-1), an \emph{intermediate}
one (GP X-2), and an \emph{external} one (GP X-3), whose
bathymetries are respectively 10 $m$, 20 $m$, 30 $m$. At Lampedusa
island samples were randomly collected along the south-eastern
part.}\medskip} \label{Palermo_Gulf}
\end{figure}
In these studies we consider only the percentage of abundances of
\emph{Adelosina} and \emph{Quinqueloculina} ($1^{st}$ group),
\emph{Elphidium} ($2^{nd}$ group) and the concentrations of heavy
metals, reported in Fig.2.
\begin{figure}[htbp]
\begin{center}
\includegraphics[width=11.7cm]{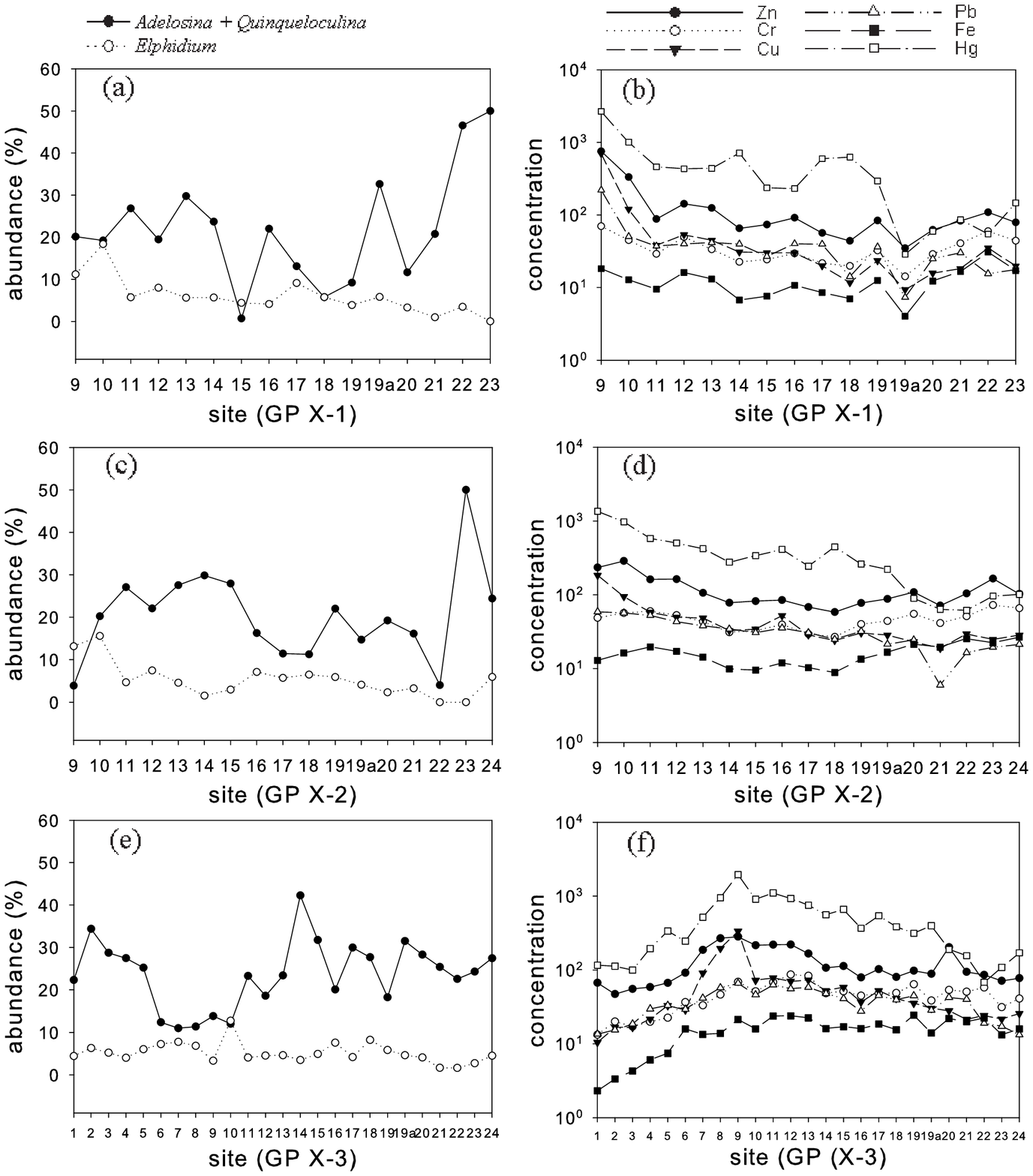}
\vskip-0.75cm
\end{center}
\caption{\small \emph{Spatial behaviour for the abundances of the
two groups of foraminifera: (a) "strip $1$", (c) "strip $2$", (e)
"strip $3$". The abundances of the two biological groups are
expressed in percentage on the total number of the foraminifera
counted in each sample. The accuracy is about 10\%. Spatial
behaviour of the concentrations of the six heavy metals: (b) "strip
$1$", (d) "strip $2$", (f) "strip $3$". The metal concentration
values are measured in $\mu gram$ per $gram$ of sediment for $Zn$,
$Cr$, $Cu$ and $Pb$, in $mgram$ per $gram$ of sediment for $Fe$, and
in $nanogram$ per $gram$ of sediment for $Hg$. The accuracy is about
$7\%$.}} \label{species_P}
\end{figure}
We note that the two groups of foraminifera show an anticorrelated
behaviour each other. Moreover the spatial behaviour of $1^{st}$ and
$2^{nd}$ groups appears respectively anticorrelated and correlated
with the metals, except two of them, namely $Cr$ and $Fe$. In order
to get a quantitative evaluation of this correlated/anticorrelated
behaviour we consider the spatial correlation coefficient given by
\begin{equation}
C=\frac{cov_{xy}}{s_x s_y} \label{corr_coeff}
\end{equation}
where
\begin{equation}
cov_{xy}=\frac{\sum_i(x_i-\bar{x})(y_i-\bar{y})}{N}
\label{covariance}
\end{equation}
is the covariance with $i$ running over the number of sites, N, of
each strip. Here $\bar{x}$, $s_x$, $\bar{y}$, $s_y$ are mean value
and root mean square, respectively of group 1 and group 2 of
foraminifera, obtained over each of the three strips.

We calculated the values of $C$ (a) between the two groups of
foraminifera and (b) between foraminifera and heavy metals. The
results are reported in Table 2. To come to conclusion about
correlations and anticorrelations we chose a threshold parameter,
$C_0$, whose values are also given in Table 2. $C_0$ was determined
by a randomization test based on some permutations of the original
series ("shuffling"). For example, by using Eq.~(\ref{corr_coeff}),
we calculated $C_r$ for a group of foraminifera, considered in the
original order of sites, and a metal in random order. By iterating
this procedure we can define the root mean squared value
\begin{equation}
C_0=\sqrt{\frac{\sum_k C_{r_k}^2}{n_{exp}}} \label{corr_coeff_mean},
\end{equation}
with $k$ running over $n_{exp}$ different experiments (numerical
realizations). For each set of data we fix the threshold at $C_0$
for correlated behaviour (meaningful correlation for $C>C_0$), and
at $-C_0$ for anticorrelated behaviour (meaningful anticorrelation
for $C<-C_0$). We performed this shuffling procedure setting
$n_{exp}=10000$. As expected, the values of $C_0$ depend on the
number of point (sites) considered, so that we obtained, within a
given strip, a threshold value $C_0$ common for all $C's$.

Metal concentrations measured in \emph{Posidonia oceanica} meadow
(Tranchina et al., 2004; Tranchina et al., 2005) and in sediments
from the Gulf of Palermo are generally higher than those measured in
sediments collected in clean sites along the Sicilian coast, and
increase considerably in the sites just near the Palermo harbour,
due to the increasing pollution (see GP 9-1, GP 9-2 and GP 9-3 in
Figs.1,2). In Fig.2 the values of $Cr$, $Cu$, $Pb$, $Zn$ are
reported in $\mu g \cdot g^{-1}$, Hg in $ng \cdot g^{-1}$, and Fe
$mg \cdot g^{-1}$.

By comparing the values of $C's$ with the corresponding $C_0's$ (see
Table 2) we observe that the abundances of \emph{Elphidium} are
correlated with the concentrations of $Cu$, $Hg$, $Pb$, $Zn$ in
strips 1 and 2, and uncorrelated in strip 3. Otherwise the group
formed by \emph{Adelosina} and \emph{Quinqueloculina} is (i)
uncorrelated with the same metals in strip 1, and anticorrelated in
strip 3, (ii) anticorrelated with $Cu$ and $Hg$ in strip 2.
Moreover, we note between the $1^{st}$ group of foraminifera and the
concentrations of Cr and Fe (i) a correlated behaviour in strips 1
and 2; (ii) an anticorrelated behaviour in strip 3. We observe that
Fe and Cr, with the concentrations we measured in our samples,
contribute to produce favourable conditions of life for benthic
foraminifera. In particular Fe represents an essential nutrient for
zooplankton. Cr$^{III}$ is not a toxic agent for the biological
species considered, since it is present in the cellular structure of
benthic foraminifera (Frausto da Silva and Williams, 2001).

The anticorrelation observed in strip 3 could be ascribed to the
contemporaneous presence of higher concentrations for the six metals
considered. In particular, comparing panels b, d, and f  Fig.2, we
note the total metal concentration increases going from strip 1 to
strip 3 when we consider the transects from GP9 to GP23. This
behaviour produces a progressive enhancement of the overall
pollution that causes a greater toxicity for biospecies. It is
therefore more interesting to focus on the global pollution produced
by the presence of all metals. In order to emphasize the fluctuating
behaviour both of the two species and six metals as a function of
the different positions (sites), for each series of data
(foraminifera and metals) we calculate the normalized abundances and
concentrations. These are obtained for each series by calculating
the mean value and normalizing the whole series respect to this
value. For the heavy metals we performed, then, in each site the
summation of all the normalized values, which we name total
normalized concentration (TNC), obtaining one series for the spatial
behaviour of the overall concentration of $Zn$, $Cu$, $Cr$, $Fe$,
$Pb$, $Hg$. The results are reported in Fig.3. Here the graphs show
more clearly the anticorrelated behaviour of the two groups of
foraminifera. Moreover some correlated or anticorrelated behaviour
between the normalized metal concentration and the two groups of
foraminifera is more visible. In order to get a quantitative
evaluation of this behaviour we calculated and reported in Table 3
$C's$ and $C_0's$ between foraminifera and metals. The values of
Table 3 confirm the presence of an anticorrelated (correlated)
behaviour between the $1^{st}$ group ($2^{nd}$ group) of
foraminifera and the total pollution for which the heavy metals are
responsible.
\begin{table}[htbp]
\begin{center}
\includegraphics[width=12cm]{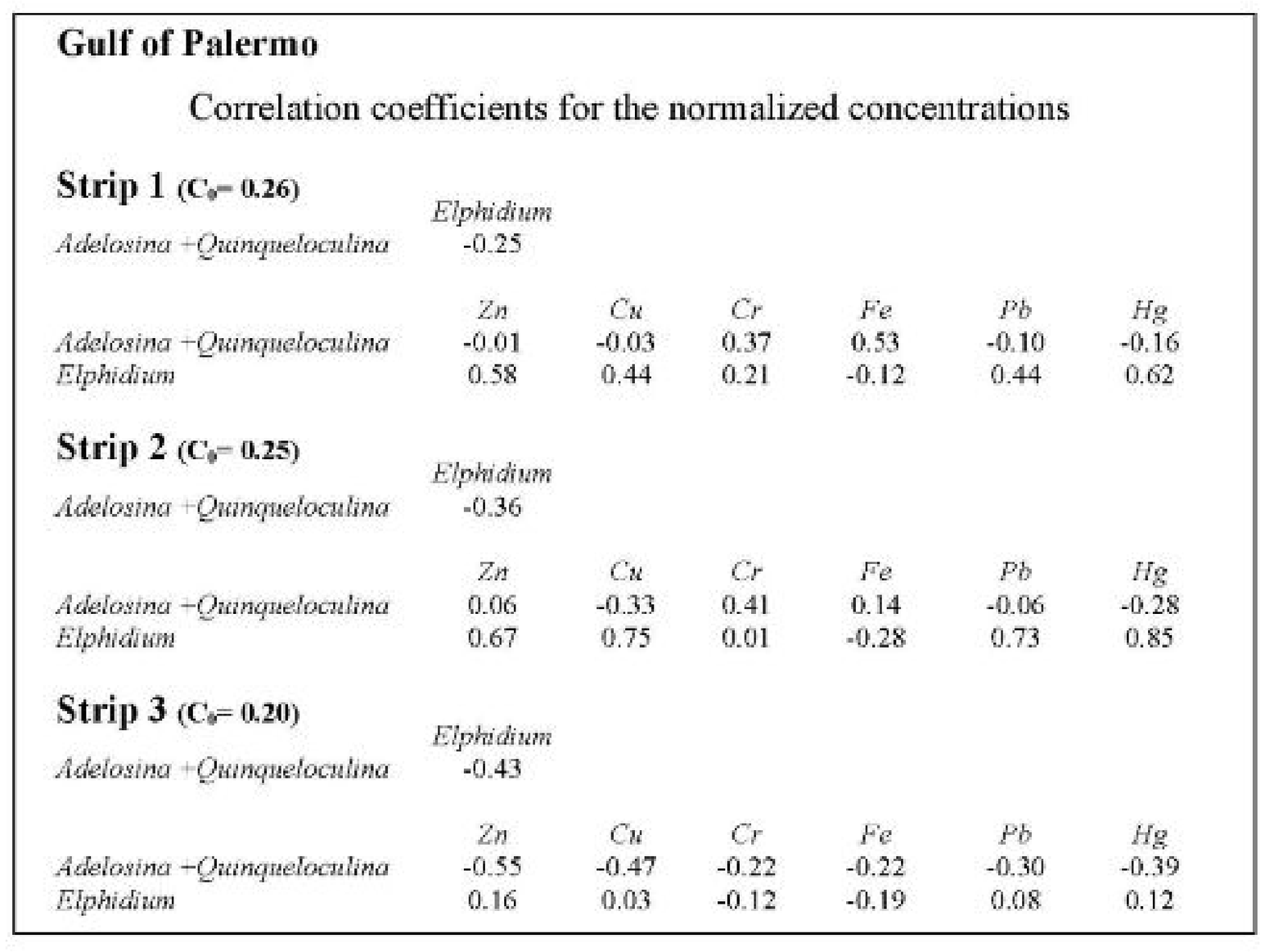}
\end{center}
\caption{ \small \emph{ Spatial correlation coefficients (i) between
the two groups of foraminifera and (ii) between foraminifera and
heavy metals.}\medskip}
 \label{table_correlation_GP_1}
\end{table}
The different responses of the two biological groups to metal
pollution can be explained by considering that some species are able
to adapt themselves to conditions that are prohibitive for other
species. In this sense, the species belonging to \emph{Elphidium}
show an opportunist behaviour,
\begin{figure}[htbp]
\begin{center}
\includegraphics[width=12cm]{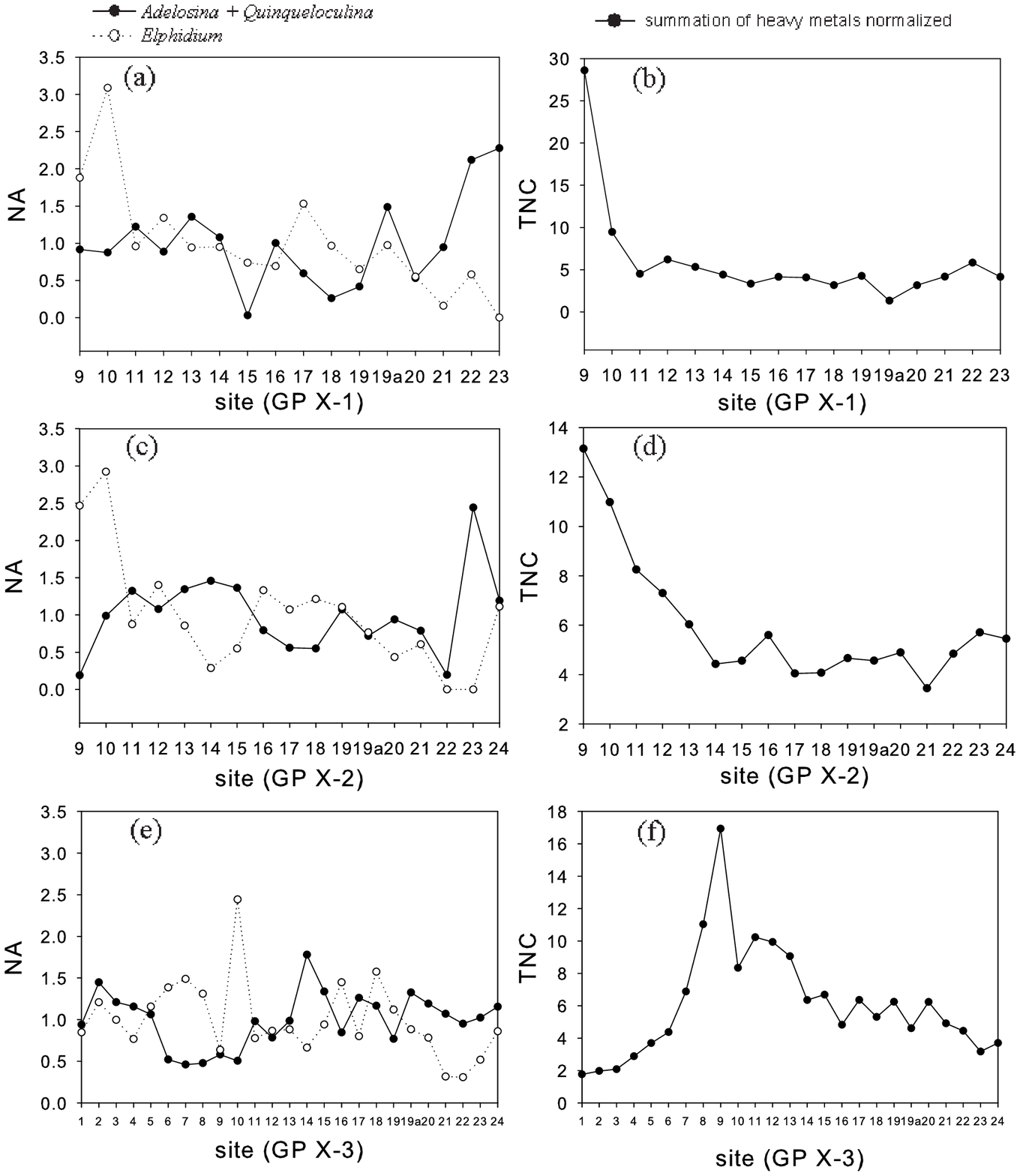}
\end{center}
\caption{ \small \emph{Spatial behaviour for the normalized
abundances (NA) of the two groups of foraminifera: (a) "strip $1$",
(c) "strip $2$", (e) "strip $3$". Spatial behaviour of total
normalized concentration (TNC) of heavy metals: (b) "strip $1$", (d)
"strip $2$", (f) "strip $3$".}\medskip} \label{speciesGP_norm}
\end{figure}
taking advantage from stressed environmental conditions. Conversely,
the species belonging to \emph{Adelosina} and \emph{Quinqueloculina}
are more sensitive to environment adverse conditions, so that in the
presence of increasing pollution their abundance decreases.
\begin{table}[htbp]
\begin{center}
\includegraphics[width=12cm]{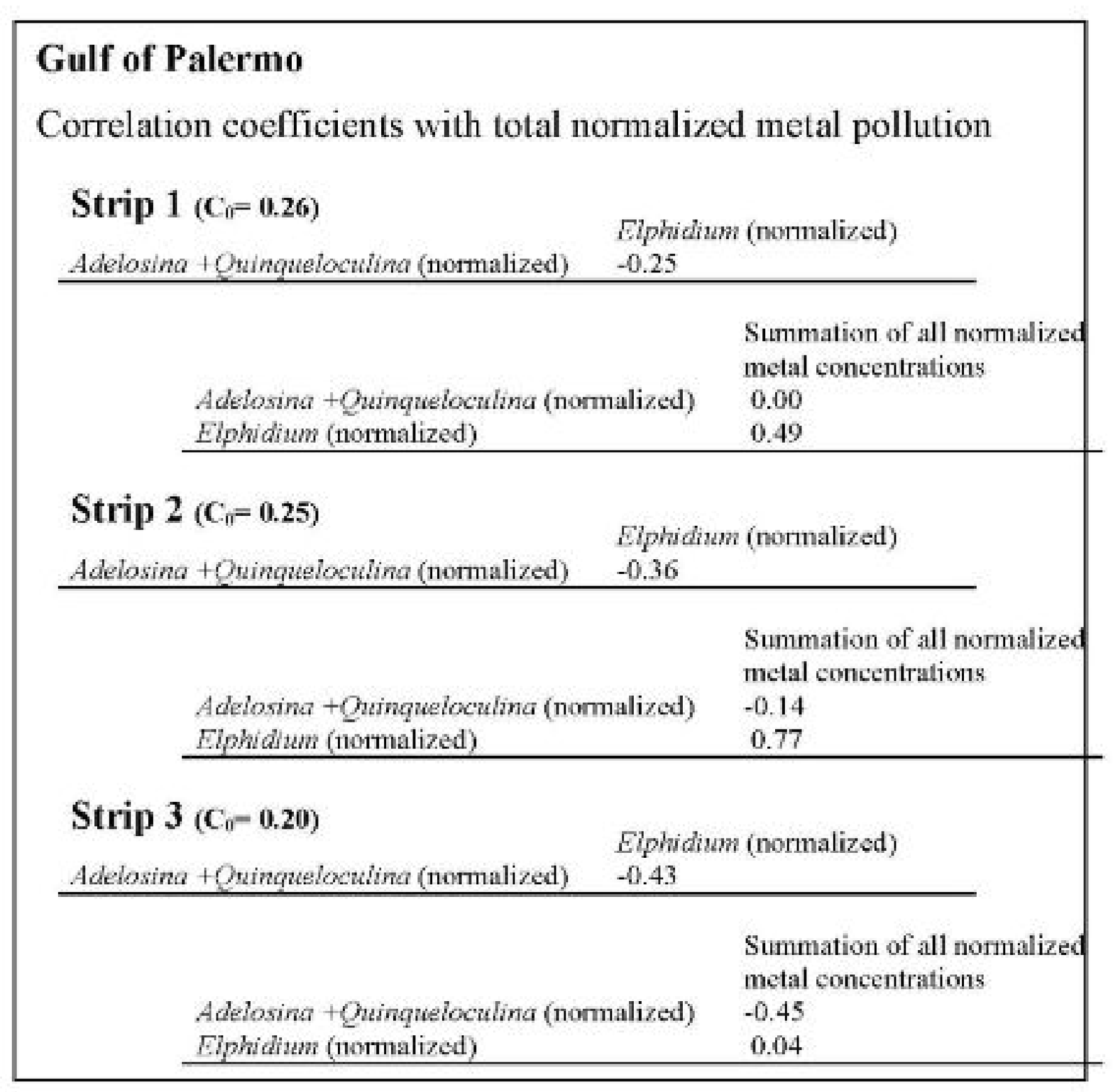}
\end{center}
\caption{ \small \emph{Spatial correlation coefficients (i) between
the two groups of foraminifera, (ii) between each group of
foraminifera and total normalized concentration of heavy
metals.}\medskip} \label{table_correlation_GP_2}
\end{table}

\subsection{Data from Lampedusa island}\label{LAMP_data}

In this paragraph we discuss experimental data obtained from $5$
samples, labeled LAMP-$6$, LAMP-$12$, LAMP-$13$, LAMP-$18$,
LAMP-$19$, that have been collected along the coast of Lampedusa
island. In view of an analysis based on a two-species competing
model (see Section~\ref{Model}), we selected the same two groups of
benthic foraminifera, already analyzed. In Fig.4 we report the
abundances of the two biological groups and the concentrations of
the six heavy metals measured in these $5$ sites. From the
inspection of the figure we note an anticorrelated behaviour between
the $1^{st}$ and $2^{nd}$ groups of foraminifera. Moreover, they
appear spatially uncorrelated and correlated with metal
concentrations, respectively.
\begin{figure}[htbp]
\begin{center}
\includegraphics[width=12cm]{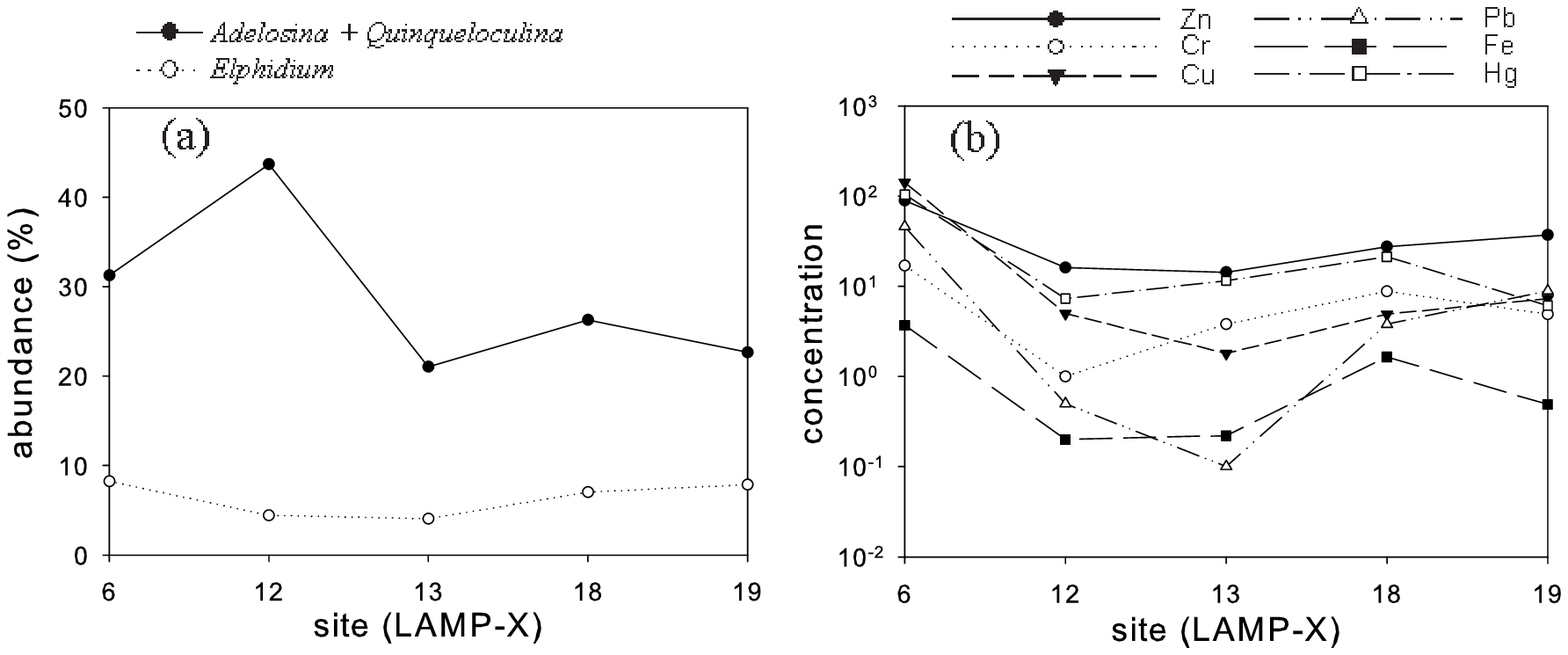}
\end{center}
\caption{ \small \emph{(a) Spatial behaviour for the abundances of
the two groups of foraminifera  in Lampedusa island; (b) Spatial
behaviour of the concentrations of the six heavy metals. Units of
measurements and accuracies are the same used in Fig.2.}\medskip}
\label{species_L}
\end{figure}
In order to get a quantitative evaluation of this behaviour we
calculated $C's$ and $C_0's$ and reported them in Table 4. Because
of the small number of sampled points the threshold value is 0.50.
This high value of $C_0$ makes statistically meaningless the values
of $C$ obtained between: (i) group 1 and group 2; (ii) group 1 and
metals. However, group 2 shows a statistically significant
correlation with metals.

This last result strengthens the trend observed in Gulf of Palermo.
\begin{table}[htbp]
\begin{center}
\includegraphics[width=12cm]{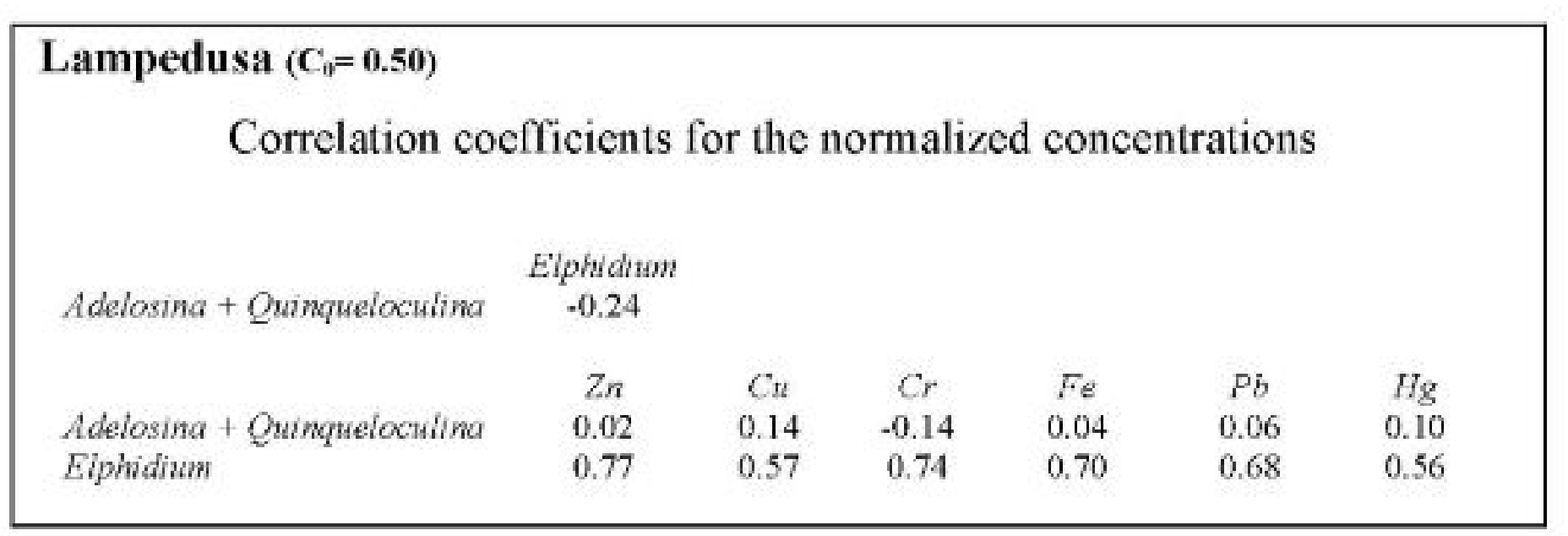}
\end{center}
\caption{ \small \emph{Spatial correlation coefficients (i) between
the two groups of foraminifera and (ii) between foraminifera and
heavy metals for data collected along the coast of
Lampedusa.}\medskip}
 \label{table_correlation_LAMP_1}
\end{table}
This uncorrelated/correlated behaviour between foraminifera and
metals becomes more evident by calculating the normalized
concentrations with the same technique of the previous section.
\begin{figure}[htbp]
\begin{center}
\includegraphics[width=12cm]{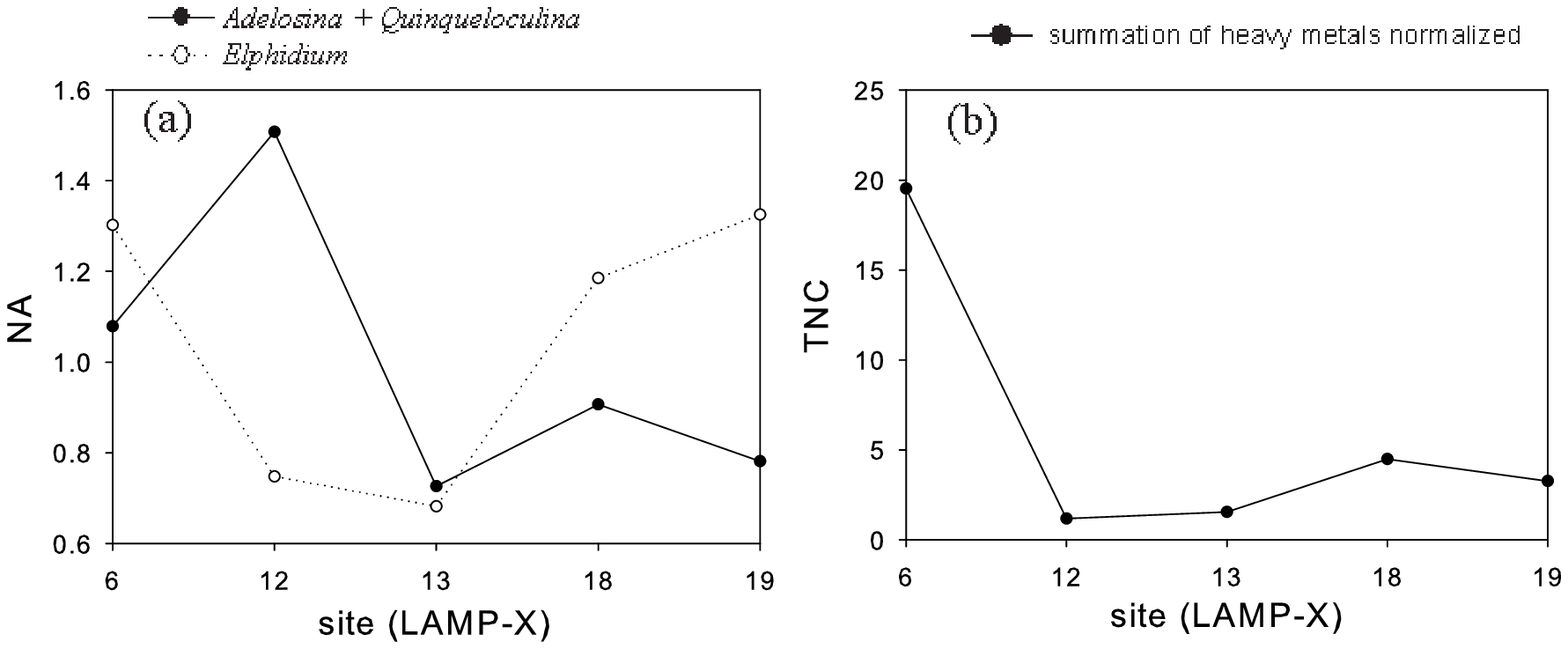}
\end{center}
\caption{ \small \emph{(a) Spatial behaviour for the normalized
abundances (NA) of the two groups of foraminifera in Lampedusa
island; (b) Spatial behaviour of total normalized concentration
(TNC) of heavy metals in Lampedusa island.}\medskip}
\label{speciesL_norm}
\end{figure}
The results are shown in Fig.5. In Table 5 we report $C's$ and
$C_0's$ between the species and between each species and total
normalized metal pollution. An inspection of Fig.5 together with the
values of Table 5 shows: (i) an almost uncorrelated behaviour
between the total normalized concentration (TNC) of metals and the
$1^{st}$ group of foraminifera, (ii) a strong correlation between
the $2^{nd}$ group of foraminifera and TNC.
\begin{table}[htbp]
\begin{center}
\includegraphics[width=12cm]{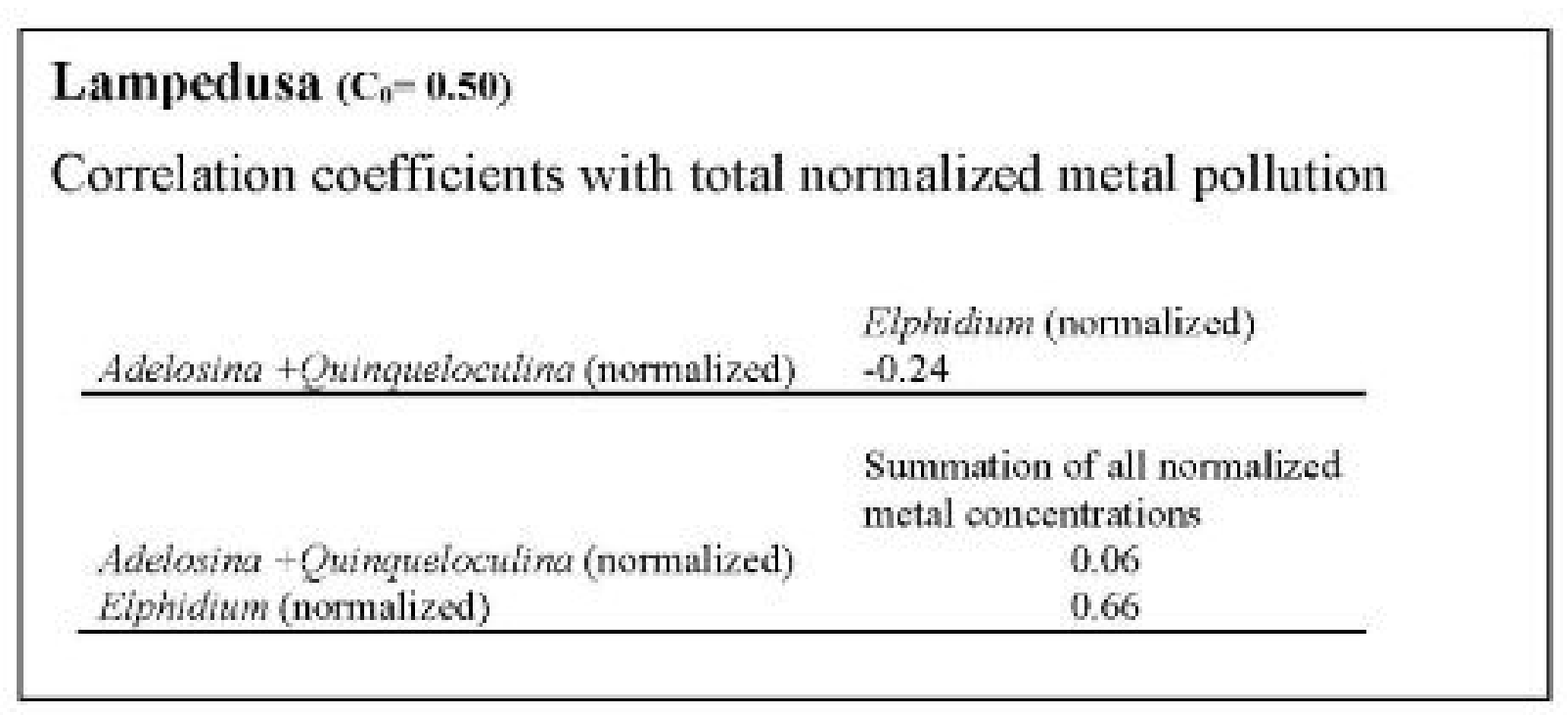}
\end{center}
\caption{ \small \emph{Spatial correlation coefficients (i) between
the two groups of foraminifera, (ii) between each group of
foraminifera and total normalized concentration of heavy metals for
data collected along the coast of Lampedusa.}\medskip}
 \label{table_correlation_LAMP_2}
\end{table}
The absence of an anticorrelation between the $1^{st}$ biological
group and the metal pollution can be explained by considering that
the global level of pollution over all the $5$ sites is quite low.
Because of this data from Lampedusa can be a good reference point
for calibration in our investigation. In fact, all the sampled
sites, except Lamp-6, which corresponds to the harbour of Lampedusa,
present low levels of metal concentrations. We note also that even
low concentrations of metals produce a correlated behaviour of the
$2^{nd}$ group of foraminifera, which show an extreme sensitivity to
the presence of metals in their habitat. The global uncorrelated
behaviour between the total normalized metal concentration and all
$1^{st}$ group of foraminifera can be considered as a signature of
clean sea waters. This can help for a better evaluation of data
coming from highly polluted waters, where a clear anticorrelation
exists between the so-called "clean" species ($1^{st}$ group) and
the metal pollution which, conversely, is well correlated with the
abundances of "opportunist" species ($2^{nd}$ group).

\section{The model}\label{Model}

In this section we present a model based on Lotka-Volterra equations
to describe the time evolution of two competing species according to
the spatial experimental data obtained for the two groups of
foraminifera. A similar model was applied to describe the time
evolution of two competing species of planktonic foraminifera
(Caruso et al., 2005).

Interaction between species and its environment is strictly
connected with the spatial behaviour of populations. A huge amount
of spatial models for ecological systems has been developed in the
last two decades (Spagnolo et al., 2004; J{\o}rgensen, 2007; Sumedha
et al., 2007). In particular among these different approaches we
mention: (i) Coupled Map Lattice (CML) model (Kaneko, 1992; Sol\'{e}
et al., 1992a; Sol\'{e} et al., 1992b; Valenti et al., 2004b);
generalized Lotka-Volterra equations with diffusive terms
(Raychaudhuri et al., 1996; Jesse, 1999; Tsekouras et al., 2004;
L\'{o}pez-S\'{a}nchez, 2005; Valenti et al., 2006); (iii)
migration-drift models based on space-time autoregressive moving
average (STARMA) processes (Epperson, 2000); (iv) simple size
structured metapopulation models based on logistic equations in the
presence of different types of dispersal among patches (Ezoea and
Nakamura, 2006); (v) approaches neglecting an explicit time
evolution, such as geostatistical models with Poisson distributions
(Monestiez et al., 2006), generalized additive models (GAM)
(Ferguson et al., 2006) and mixed regressive-spatial autoregressive
models (Overmars et al., 2003); (vi) probability transition models
with random dispersal (Zhang et al., 2006); (vii) predictive
vegetation models based on statistical methods (Miller et al.,
2007). However, all these models account for spatial behaviour by
considering spatial structures of data and fitting procedures and/or
diffusive phenomena. Conversely, in our model we neglect the
transport effects due to marine currents and water flows, since
benthic foraminifera lie in seabed. In this context transport of
biospecies could occur because some individuals move from a highly
populated site towards another one, where life conditions are more
favourable. However, foraminifera are not able to change their
positions, so that they are confined inside a given site. Therefore
we choose a time evolution model, without considering the spatial
dynamics. Moreover, in order to take into account the effects of
heavy metal pollution and its diffusion, we consider that the metal
concentrations, affecting the foraminifera abundances, are not
constant over time but undergo random fluctuations. This behaviour
can be also connected with passive diffusive processes from the
sediment to the water column and viceversa. Therefore, in view of
devising a predictive dynamical model we consider metal pollution as
a noise source, so that the biospecies are governed by a stochastic
dynamics. As we saw in the previous section, from experimental data
we found that, through the different sampling sites, the abundances
of the two biological groups show a fluctuating behaviour in which
we were able to recognize some peculiarities. In particular we note
that the first group (\emph{Adelosina} and \emph{Quinqueloculina})
appears to be more sensitive to the environmental pollution
(presence of heavy metals) with regards to their survival, playing a
role of sentinel species (Jamil, 2001). The second one
(\emph{Elphidium}), differently, survives in the presence of
environmental pollution due to the metal concentrations. We can
consider, therefore, the pollution as an environmental disturbance,
whose intensity is given by the value of its concentration. It is
then reasonable to think of the two populations as two competing
species in the presence of an environmental noisy pollution. It is
important to note that the competitive behaviour of the two
populations is mediated by several environmental parameters, as
water dissolved oxygen, sediment grain-size composition, presence of
seagrass in the soft-bottom, etc. The heavy metal pollution plays
the role of an external perturbation of the normal life conditions,
by modifying, for example, the growth rates of the two populations.
Moreover we note that in a dynamical model of an ecological system
it is reasonable to consider that the values of the environmental
parameters vary during the time evolution. In particular the
pollution represents a parameter subject to fluctuations whose
behaviour is difficult to predict, so that its concentration at a
given time can be considered as a random variable, that is a noise.
Therefore, to model the dynamics of the two groups of foraminifera
we use the generalized Lotka-Volterra equation
\begin{eqnarray}
\frac{dx}{dt}= x\thinspace(\alpha_x-x-\beta_x y)+x\thinspace\xi_x(t)\label{LotVolx}\\
\frac{dy}{dt}= y\thinspace(\alpha_y-y-\beta_y
x)+y\thinspace\xi_y(t),
 \label{LotVoly}
\end{eqnarray}
in the presence of a multiplicative noise, which mimics the random
fluctuations of the total normalized metal concentration (TNC). In
Eqs.~(\ref{LotVolx}) and (\ref{LotVoly}), $\alpha_x$ and $\alpha_y$
are the growth rates of species $x$ and species $y$, $\beta_x$ and
$\beta_y$ the coefficients of the interspecies competition. Here
$\xi_x(t)$ and $\xi_y(t)$ are defined respectively as non-positive
and non-negative white noises
\begin{eqnarray}
\xi_x(t) &=& -\vert \xi(t) \vert\label{noisex}\\
\xi_y(t) &=& +\vert \xi(t) \vert,
 \label{noisey}
\end{eqnarray}
where $\xi(t)$ is a Gaussian white noise with the usual statistical
properites
\begin{equation}
\langle \xi(t)\rangle = \; 0, \;\;\;\;\;\;\; \langle
\xi(t)\xi(t')\rangle = \; \sigma \delta(t-t') \label{white}
\end{equation}
and $\vert \xi(t) \vert$ indicates the absolute value of $\xi(t)$.
The semi-Gaussian noise sources allow to take into account the
different effects that the random fluctuations of heavy metal
concentrations produce in the dynamics of the two foraminifera
groups. TNC appears to be spatially anticorrelated with the $1^{st}$
group of foraminifera and correlated with the $2^{nd}$ one. In
Eqs.~(\ref{LotVolx}),~(\ref{LotVoly}) $\xi_x(t)$ and $\xi_y(t)$
produce respectively a decrease and an enhancement in the species
densities. Therefore the noise reduces the "natural" growth of one
population and helps that of the other one. Concerning the
deterministic dynamics, that is in the absence of noise, the
stationary states of the species concentrations in the coexistence
regime are
\begin{eqnarray}
x_{st} &=& \frac{\alpha_x-\beta_x \thinspace
\alpha_y}{1-\beta_x\thinspace \beta_y}, \label{stationary_x}\\
y_{st} &=& \frac{\alpha_y-\beta_y \thinspace
\alpha_x}{1-\beta_x\thinspace \beta_y}, \label{stationary_y}
\end{eqnarray}
and the following conditions
\begin{eqnarray}
\alpha_x &>& \alpha_y \beta_x, \;\;\;\;\; \alpha_y > \alpha_x \beta_y \label{coexistence1}\\
\beta_x \beta_y &<& 1,
 \label{coexistence2}
\end{eqnarray}
should be satisfied in such a way that both species survive.
Otherwise an exclusion regime is established, that is one of the two
species vanishes after a certain time. Coexistence and exclusion of
one of the two species correspond to stable states of the
Lotka-Volterra's deterministic model (Bazykin, 1998).

\subsection{Theoretical results}\label{theoretical_results}

 To describe the dynamics of our real system, where the two groups of
foraminifera are contemporary present, we assume that the two
populations are in coexistence regime, choosing the values of
$\alpha_x$, $\beta_x$, $\alpha_y$, $\beta_y$ according to the
inequalities~(\ref{coexistence1}) and (\ref{coexistence2}). We
obtain the time series of the two species for different levels of
the noise intensity $\sigma_x$ and $\sigma_y$, whose values are
obtained from TNC (see Fig.3b,d,f). For each value of TNC, that we
indicate with the symbol $P$, we calculate the decimal logarithm of
the quantity $(1 + P)$ to obtain positive values for the noise
intensity, and multiply it by a scale factor ($f_x$ for species $x$
and $f_y$ for species $y$)
\begin{eqnarray}
\sigma_x &=& f_x \thinspace Log (1+P)\label{noise_intensities1} \\
\sigma_y &=& f_y \thinspace Log (1+P).
 \label{noise_intensities2}
\end{eqnarray}
The use of the logarithm allows to reduce the fluctuations of
experimental data. The scaling factor is used to calibrate the noise
intensity with respect to the value chosen for the interaction
parameter $\beta$. So that we control the effect of the
environmental interaction and the interspecies competition. Here we
consider the same value for $f_x$ and $f_y$
\begin{equation}
\sigma = \sigma_x = \sigma_y = f \thinspace Log (1+P),
\label{noiseintensity}
\end{equation}
obtaining for both species the same value of noise intensity
$\sigma$ for each value of the TNC.

\subsection{Time series and spatial distributions}\label{Spatial_GP}

First we investigate the effect of the noise on the time behaviour
of the species. Since the dynamics of the species strongly depends
on the value of the multiplicative noise intensity, we initially
analyze the time evolution of $x(t)$ and $y(t)$ for different levels
of the multiplicative noise. In order to obtain the densities of the
two species at different points (sites) we integrate
Eqs.~(\ref{LotVolx}) and (\ref{LotVoly}) by setting
$\sigma=\sigma_x=\sigma_y$ at different values, which correspond to
TNC's measured in our samples. By following this procedure, and
taking into account the scaling of Eqs.~(\ref{noise_intensities1})
and (\ref{noise_intensities2}), we obtain the time series of the two
species for different levels of the pollution. The time series for
the two populations are obtained by setting $\alpha_x=3.0$,
$\alpha_y=2.0$, $\beta_x=7.5 \cdot 10^{-1}$, $\beta_y=3.3 \cdot
10^{-1}$, $f=5.5 \cdot 10^{-5}$. We applied a trial and error
procedure to select the values of the parameters $\alpha_x$,
$\alpha_y$, $\beta_x$, $\beta_y$, $f$, for which we obtain the best
fitting between theoretical results and experimental data. The
initial values for the two species densities are $x(0) = y(0) = 1$.
In Fig.6 we report the time series for $\sigma=0$, (TNC $= 0$, no
pollution), $\sigma=2.01 \cdot 10^{-6}$ (TNC $= 1.32$), which is the
lowest level of pollution measured, in the site GP~$19$a-$1$, and
$\sigma =8.09 \cdot 10^{-6}$ (TNC $= 28.60$), which is the highest
level of pollution, measured in the site GP~$9$-$1$.
\begin{figure}[htbp]
\begin{center}
\includegraphics[width=9cm]{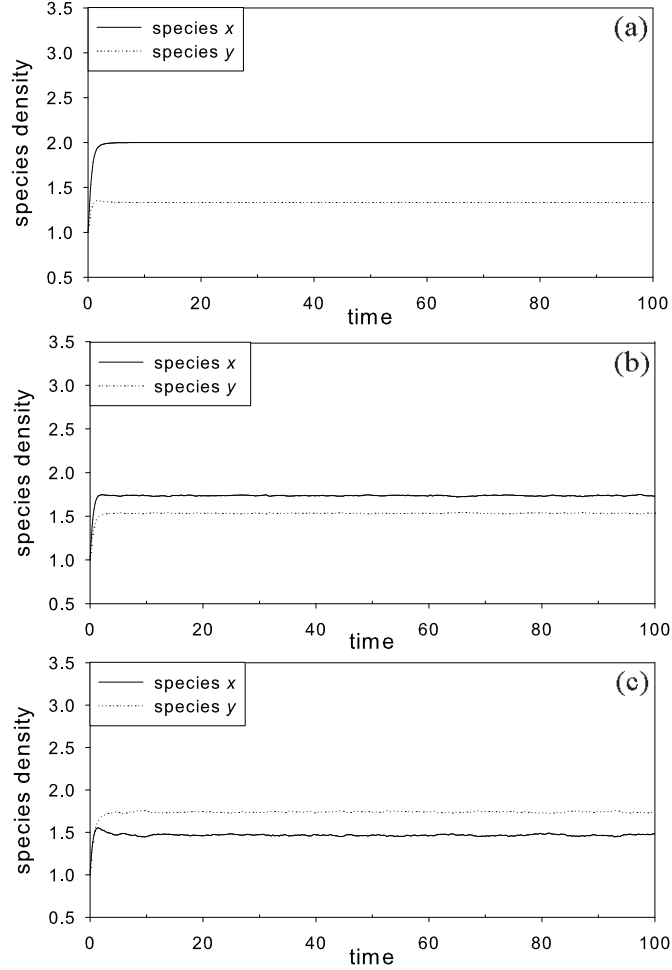}
\end{center}
\caption{ \small \emph{Time evolution of the two species for
different values of the total normalized metal concentration (TNC):
(a) TNC = 0 ($\sigma = 0$); (b) TNC = 1.32 ($\sigma =2.01 \cdot
10^{-6}$); (c) TNC = 28.60 ($\sigma =8.09 \cdot 10^{-6}$). Both time
and species densities are expressed in arbitrary units (a. u.). The
values of the parameters are $\alpha_x=3.0$, $\alpha_y=2.0$,
$\beta_x=7.5 \cdot 10^{-1}$, $\beta_y=3.3 \cdot 10^{-1}$, $f=5.5
\cdot 10^{-5}$. The initial conditions are x(0) = y(0) =
1.}\medskip}
 \label{time_series_GP}
\end{figure}
In Fig.6 we note that, after a short transient, which is not clearly
visible because of the scale used for the horizontal axis, both
species reach the stationary values that depend on the
multiplicative noise intensity $\sigma$. In the deterministic regime
these values are given by $x_{st}$, $y_{st}$ of
Eqs.~(\ref{stationary_x}) and (\ref{stationary_y}). In the presence
of semi-Gaussian noise sources these values undergo variations, i.e.
decrease for species $x$ (non-positive noise), and increase for
species $y$ (non-negative noise). In the absence of noise (Fig.6a)
species $x$ prevails, reaching a stationary value of density quite
bigger than species $y$. For a low level of noise (Fig.6b) the
stationary values for the densities of species $x$ and species $y$
decreases and increases respectively. For higher noise intensities
(Fig.6c) an inversion of population densities occurs and the density
of species $y$ exceeds that of species $x$.

In order to model the spatial distributions obtained from
experimental data for the two groups of foraminifera, for each value
of TNC we integrate numerically
Eqs.~(\ref{LotVolx}),~(\ref{LotVoly}), averaging on $1000$
simulative experiments. Then, for every noise intensity which
corresponds to a given pollution level (according to
Eq.~(\ref{noiseintensity})), we take the stationary values for both
species. By this procedure, for each species we get $58$ abundance
values. The results are reported in Fig.7.
\begin{figure}[htbp]
\begin{center}
\includegraphics[width=12cm]{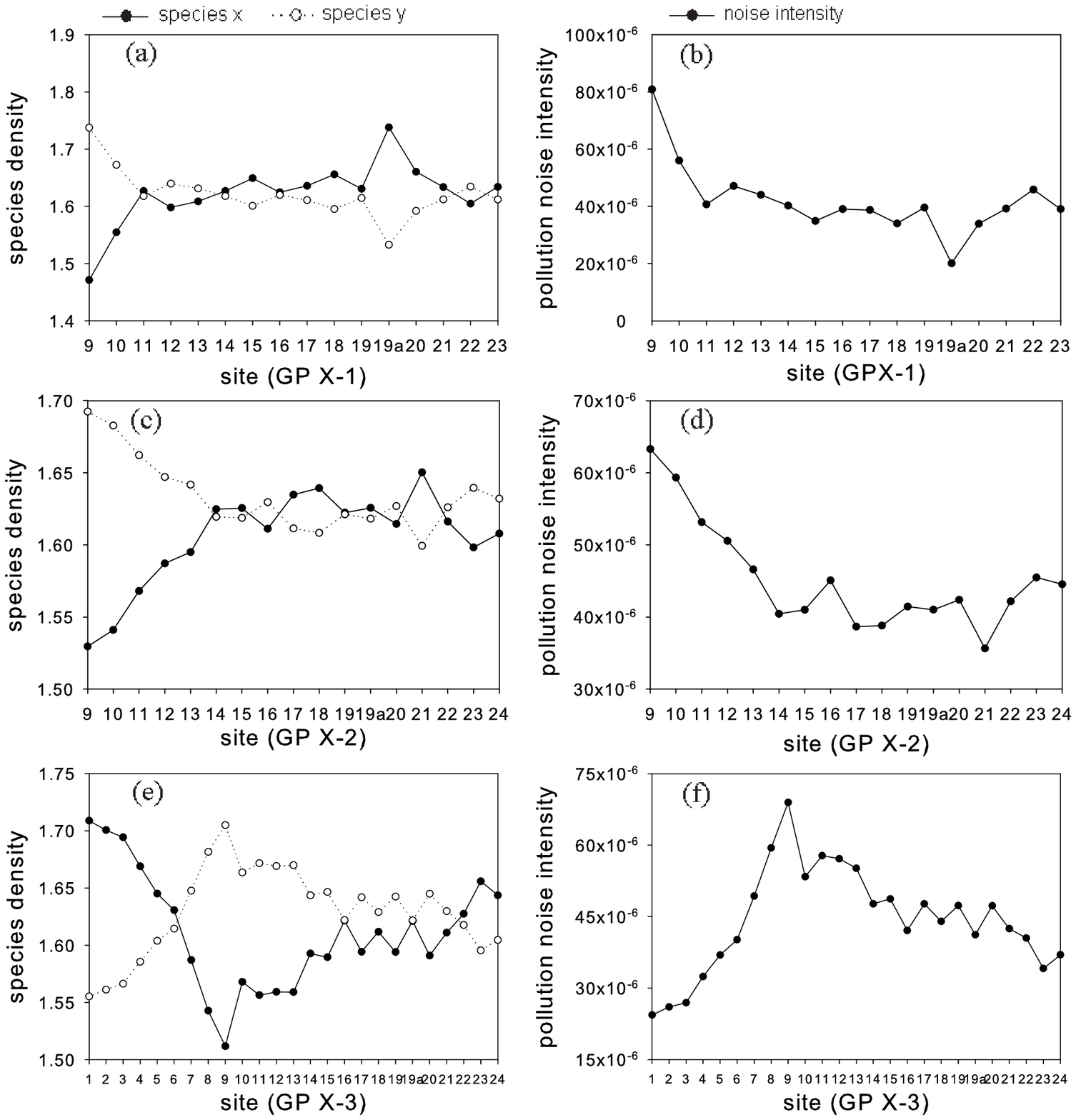}
\end{center}
\caption{ \small \emph{Theoretical spatial behaviour for the
densities of the two species in the Gulf of Palermo obtained by
numerical integration of Eqs.~(\ref{LotVolx}),~(\ref{LotVoly}) and
using, as pollution noise intensity, $\sigma=f \thinspace Log (1+P)$
where $P$ is the TNC: (a) strip $1$, (c) strip $2$, (e) strip $3$.
We set $f=5.5\cdot10^{-5}$. Spatial behaviour for the pollution
noise intensity $\sigma$: (b) strip $1$, (d) strip $2$, (f) strip
$3$. Both species densities and concentrations are expressed in
arbitrary units (a. u.). Values of parameters and initial conditions
are the same of Fig.6. The values of the species densities at each
site are the stationary values obtained from
Eqs.~(\ref{LotVolx}),~(\ref{LotVoly}).}\medskip}
 \label{spatialGP}
\end{figure}
In particular, in Fig.7b,d,f we report the values of $\sigma$
calculated from real data. We let the Lotka-Volterra system to
evolve until the stationary regime is reached (see Fig.6).
Afterwards, for the two species densities we consider the stationary
values and we report them in Fig.7a,c,e. As a consequence of the
model used the behaviour is (i) anticorrelated between the two
species, (ii) anticorrelated between the species $x$ and the
pollution noise intensity, (iii) correlated between the species $y$
and the pollution noise intensity (see Table 3). However, the choice
of the parameter values allows to obtain a good agreement between
theoretical results and experimental data (compare Fig.7 with
Fig.3).
\begin{figure}[htbp]
\begin{center}
\includegraphics[width=12cm]{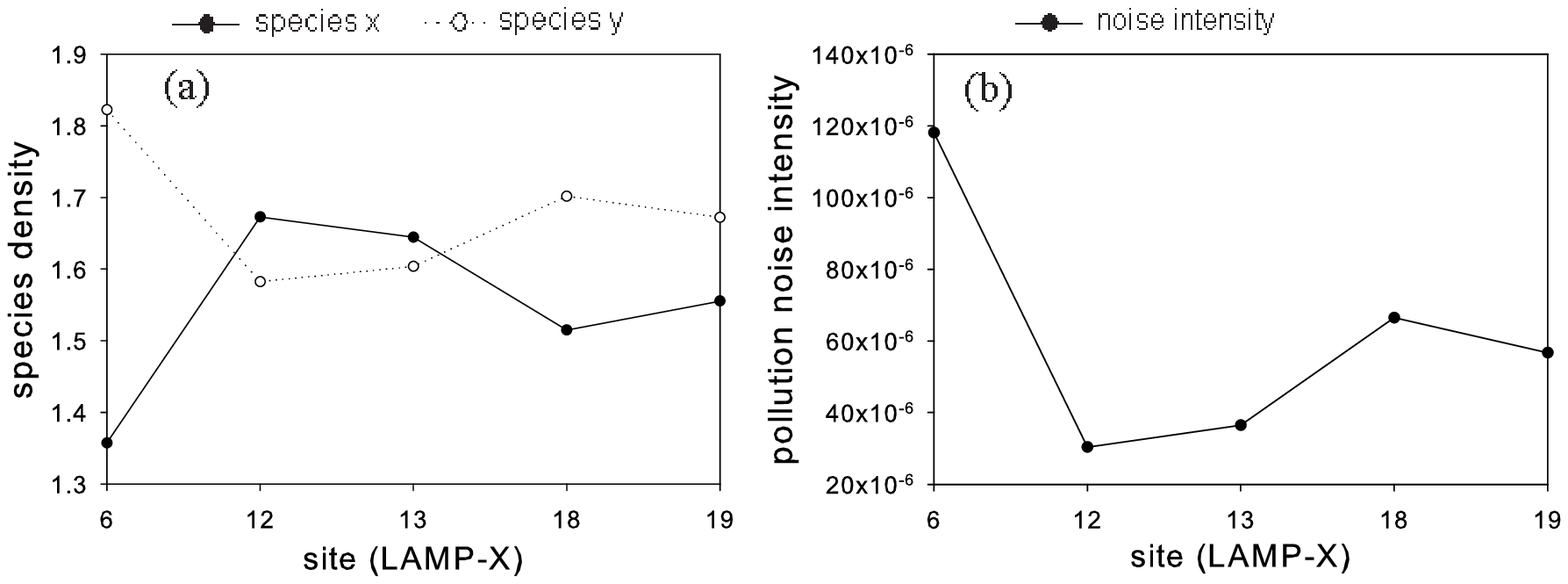}
\end{center}
\caption{ \small \emph{(a) Theoretical spatial behaviour for the
densities of the two theoretical species obtained by numerical
integration of Eqs.~(\ref{LotVolx}),~(\ref{LotVoly}) and using, as
noise intensity, $\sigma=f \thinspace Log (1+P)$ where $P$ is the
TNC in Lampedusa island. We set $f=9.0\cdot10^{-5}$. (b) Spatial
behaviour for the pollution noise intensity in Lampedusa island.
Both species densities and concentrations are expressed in arbitrary
units (a. u.). Values of parameters and initial conditions are the
same of Fig.6. The values of the species densities at each site are
the stationary values obtained from
Eqs.~(\ref{LotVolx}),~(\ref{LotVoly}).}\medskip}
 \label{spatialL}
\end{figure}
Finally, as we did for the data sampled in Gulf of Palermo, we
applied the generalized Lotka-Volterra system of
Eqs.~(\ref{LotVolx}),~(\ref{LotVoly}) to reproduce the spatial
behaviour obtained for the two biological groups along the coast of
Lampedusa island. For each species we get $5$ abundance values and
we report them in Fig.8a. Moreover we present in Fig.8b the
behaviour of $\sigma$ calculated from real data. The behaviour
reported in Fig.8a shows a good qualitative agreement with the
experimental data (compare Fig.5 with Fig.8a). In particular, from
Fig.8a we note a spatial anticorrelation between the two theoretical
species and a correlation between species $y$ (which mimics the
$2^{nd}$ group of foraminifera) and the noise intensity (see Table
5). We note that the values of $f$ used to obtain the theoretical
results of Figs.~7,~8 are different, because of the different
fluctuating spatial behaviour of TNC in the two Sicily areas.

\section{Conclusions}

In this paper we studied the spatial distributions of some species
of zoobenthos and heavy metals, taking into account their
correlations. In particular we focused on two groups of benthic
foraminifera, the first one constituted of \emph{Adelosina} spp. and
\emph{Quinqueloculina} spp., the second one formed by
\emph{Elphidium} spp. From the data analysis we found a spatial
anticorrelation between the two biological groups. Moreover we
observed an anticorrelated behaviour between the $1^{st}$ group of
foraminifera and the overall metal pollution (TNC) in strip 3 where
TNC values show marked fluctuations. Conversely, the $2^{nd}$ group
shows an "opportunistic" behaviour, increasing its abundance in the
presence of higher levels of TNC. We modeled the behaviour of these
two foraminiferal groups, by introducing a generalized
Lotka-Volterra (LV) system with multiplicative Gaussian noise which
takes into account TNC effect. The theoretical spatial behaviour
obtained by our model agrees qualitatively with that obtained from
the experimental data, where the $1^{st}$ and $2^{nd}$ groups of
foraminifera show respectively anticorrelated and correlated
behaviour with TNC. Benthic foraminifera abundances and metal
concentrations from Lampedusa island show a behaviour similar to
that found for the Gulf of Palermo, but with a peculiarity: the
uncorrelated spatial behaviour between the $1^{st}$ group of
foraminifera and TNC. This specific spatial behaviour could be
considered as a signature of clean sea water sites. The theoretical
results, obtained with the same LV system, are in a good qualitative
agreement with the experimental data. We note finally that the model
proposed should be useful to explain the time evolution of
ecological species, whose dynamics is strongly affected by
fluctuations of environmental parameters (Zimmer, 1999; Bjornstad
and Grenfell, 2001; Caruso et al., 2005), including temperature,
food availability and, as in the above analyzed case, pollution.

\section{Acknowledgments}
This work has been supported by ARPA Sicilia and Universit\`a degli
Studi di Palermo. We wish to thank Dr. Mario Badami for his
technical support with his sailing boat during the sampling of the
sites in the Gulf of Palermo, and Dr. G. Augello for useful
discussions about numerical simulations.

\section*{References}

\begin{description}

\item \qquad Bazykin, A.D., 1998. Nonlinear dynamics of interacting populations, World Scientific Series on Nonlinear
Science, Series A, Vol. 11, Singapore.

\item \qquad Bj{\o}rnstad, O. N., Grenfell, B. T., 2001. Noisy Clockwork:
Time Series Analysis of Population Fluctuations in Animals. Science
293, 638-643.

\item \qquad Blake, W. J., Kaern, M., Cantor, C. R., Collins, J. J., 2003. Noise in
eukaryotic gene expression. Nature 422, 633-637.

\item \qquad Blasius, B., Huppert, A., Stone, L., 1999. Complex dynamics and
phase synchronization in spatially extended ecological systems.
Nature 399, 354-359.

\item \qquad Brown J. H., Whitham, T. G., Ernest, S. K. M., Gehring, C. A., 2001.
Complex Species Interactions and the Dynamics of Ecological Systems:
Long-Term Experiments. Science 293, 643-650.

\item \qquad Caruso, A., 2004. Climatic Changes During Upper Pliocene/Lower
Pleistocene at Capo Rossello (Sicily, Italy): Response from
Planktonic Foraminifera Approach. In Coccioni et al. eds. Special
Volume Grzybowski Foundation, University of London 9, 17-36.

\item \qquad Caruso, A., Gargano, M. E., Valenti, D., Fiasconaro, A., Spagnolo,
B., 2005. Climatic Changes and Role of Noise in Planktonic
Foraminifera in the Mediterranean Sea. Fluc. Noise Lett. 5,
L349-L355.

\item \qquad Chichigina, O., Valenti, D. and Spagnolo, B., 2005.
A Simple Noise Model with Memory for Biological Systems, Fluc. Noise
Lett., 5, L243-L250.

\item \qquad Cirone, M. A., de Pasquale, F., Spagnolo, B., 2003. Nonlinear
relaxation in population dynamics. Fractals 11, 217-226.

\item \qquad Ciuchi, S., de Pasquale, F., Spagnolo, B., 1996. Self-regulation
mechanism of an ecosystem in a non-Gaussian fluctuation regime.
Phys. Rev. E 54, 706-716; \emph{ibidem}, 1993. Non linear Relaxation
in the presence of an Absorbing Barrier. Phys. Rev. E 47, 3915-3926.

\item \qquad Cook, J. M., Gardner, M. J., Griffiths, A. H., Jessep, M. A.,
Ravenscroft, J. E., Yates, R., 1997. The comparability of sample
digestion techniques for the determination of metals in sediment.
Marine Pollution Bulletin 34, 637-644.

\item \qquad Epperson, B. K., 2000. Spatial and Space–Time Correlations in
Ecological Models. Ecol. Model. 132, 63–76.

Ezoea, H., Nakamura, S., 2006. Size Distribution and Spatial
Autocorrelation of Subpopulations in a Size Structured
Metapopulation Model. Ecol. Model. 198, 293–300.

\item \qquad Ferguson, M.C., Barlowa, J., Fiedler, P., Reilly, S.B., Gerrodette,
T., 2006. Spatial models of delphinid (family Delphinidae) encounter
rate and group size in the eastern tropical Pacific Ocean. Ecol.
Model. 193, 645-662.

\item \qquad Frausto da Silva, J. J. R., Williams, R. J. P., 2001.
The Biological Chemistry of the Elements: The Inorganic Chemistry of
Life. Oxford University Press USA.

\item \qquad Freund, J. A., Pöschel, T. (Eds.), 2000. Stochastic Processes in
Physics, Chemistry, and Biology, Lecture Notes in Physics, vol. 557,
Springer, Berlin

\item \qquad Giardina, I., Bouchaud, J. P., Mezard, M., 2001. Proliferation
Assisted Transport in a Random Environment. J. Phys. A: Math. Gen
34, L245-L252.

\item \qquad Gielen, J. L. W., 2000. A Stochastic Model for Epidemics Based on the
Renewal Equation. J. Biological Systems 8, 1-20.

\item \qquad Goldenfeld, N., Kadanoff,  L.P., 1999. Simple Lessons from
Complexity. Science 284, 87-89.

\item \qquad  Jamil, K. 2001. Bioindicators and biomarkers of environmental
pollution and risk assessment. Science Publishers, Enfield, New
Hampshire, USA.

\item \qquad Jesse, K.J., 1999. Modelling of a diffusive
Lotka-Volterra-System: the climate-induced shifting of tundra and
forest realms in North-America. Ecol. Model. 123, 53-64.

\item \qquad J{\o}rgensen, S.E., 2007. Two hundred volumes of
Ecological Modelling. Ecol. Model. 200, 277-278.

\item \qquad Kaneko, K., 1992. Overview of Coupled Map Lattices. Chaos 2, 279-282.

\item \qquad King, A. A., Schaffer, W. M., 2001. The Geometry of a Population Cycle:
A Mechanistic Model of the Snowshoe Hare Cycle. Ecology 82, 814-830.

\item \qquad La Barbera, A., Spagnolo, B., 2002. Spatio-temporal patterns in
population dynamics. Phys. A 314, 120-124.

\item \qquad Loeblich, A. R., Tappan, Jr., 1964. H. Sarcodina, chiefly 'Thecamoebians'
and Foraminifera. In: Moore, R. C. (Ed.), Treatise on Invertebrate
Paleontology. University of Kansas Press, New York.

\item \qquad Loeblich, A. R., Tappan, Jr. H., 1988. Foraminiferal Genera and their Classification
vol. 4. Van Nostrand Reinhold, New York, 970 pp.

\item \qquad L\'{o}pez-S\'{a}nchez, J.F., Alhama, F.,
Gonz\'{a}lez-Fern\'{a}dez, C.F., 2005. Introduction and permanence
of species in a diffusive Lotka-Volterra system with time-dependent
coefficients. Ecol. Model. 183, 1-9.

\item \qquad Lourens, L. J., Antonarakou, A., Hilgen, F. J., Van Hoof, A. A. M., Vergnaud Grazzini, C.,
Zachariasse, W. J., 1996. Evaluation of the Pliocene to Early
Pleistocene Astronomical Time Scale, Paleoceanography 11, 391-413.

\item \qquad Man, K. W., Zheng, J., Leung, A. P. K., Lam, P. K. S., Lam, M. H. W., Yen,
Y. F., 2004. Distribution and behavior of trace metals in the
sediment and porewater of a tropical coastal wetland. Science of the
Total Environment 327, 295-314.

\item \qquad Manta, D. S., Angelone, M., Bellanca, A., Neri, R., Sprovieri, M.,
2002. Heavy metals in urban soils: a case study from the city of
Palermo (Sicily), Italy. The Science of the Total Environment 300,
229-243.

\item \qquad Miller, J., Franklin, J., Aspinall, J., 2007. Incorporating spatial
dependence in predictive vegetation models. Ecol. Model. 202,
225–242.

\item \qquad Monestieza, P., Dubrocab, L., Bonninc, E., Durbecc, J.-P., Guinetb,
C., 2006. Geostatistical modelling of spatial distribution of
\emph{Balaenoptera physalus} in the Northwestern Mediterranean Sea
from sparse count data and heterogeneous observation efforts. Ecol.
Model. 193, 615-628.

\item \qquad Overmars, K.P., de Koning, G.H.J., Veldkamp, A., 2003. Spatial
autocorrelation in multi-scale land use models. Ecol. Model. 164,
257–270.

\item \qquad Ozbudak, E. M., Thattai, M., Kurtser, I., Grossman, A. D., van
Oudenaarden, A., 2002. Regulation of noise in the expression of a
single gene. Nat. Genet. 1, 69-73.

\item \qquad Raychaudhuri, S., Sinha, D.K., Chattopadhyay, J., 1996.
E ffect of time-varying cross-diffusivity in a two-species
Lotka-Volterra competitive system. Ecol. Model. 92, 55-64.

\item \qquad Rozenfeld, A. F., Tessone, C. J., Albano, E., Wio, H. S., 2001. On the
influence of noise on the critical and oscillatory behavior of a
predator–prey model: Coherent stochastic resonance at the proper
frequency of the system. Phys. Lett. A 280, 45-52.

\item \qquad Sen Gupta, B. K., 2003. Foraminifera in Marginal Marine Environments.
In: Sen Gupta, B. K. (Ed.), Modern Foraminifera. Kluwer Academic
Publishers, New York, 141-159 pp.

\item \qquad Sgarrella, F., Montcharmont Zei, M., 1993. Benthic Foraminifera of
the Gulf of Naples (Italy): systematics and autoecology. Bollettino
della Societ\`{a} Paleontologica Italiana, 32, 2, 1145-264.

\item \qquad Scheffer, M., Carpenter, S., Foley, J. A., Folke, C., Walker, B.,
2001. Catastrophic shifts in ecosystems. Nature 413, 591-596.

\item \qquad Sol\'{e}, R.V., Bascompte, J., Valls, 1992a.
Nonequilibrium Dynamics in Lattices Ecosystems: Chaotic stability
and dissipative structures. Chaos 2, 387-395.

\item \qquad Sol\'{e}, R.V., Bascompte, J., Valls, 1992b. Stability and Complexity
of Spatially Extended Two-Species Competition. J. Theor. Biol., 159,
469-480.

\item \qquad Spagnolo, B., Cirone, M., La Barbera, A., de Pasquale, F., 2002a.
Noise Induced Effects in Population Dynamics. J. Phys.: Cond. Matter
14, 2247-2255.

\item \qquad Spagnolo, B., La Barbera, A., 2002b. Role of the noise on the
transient dynamics of an ecosystem of interacting species. Phys. A
315, 114-124.

\item \qquad Spagnolo B., Fiasconaro, A., Valenti, D., 2003. Noise Induced
Phenomena in Lotka-Volterra Systems. Fluc. Noise Lett. 3, L177-L185.

\item \qquad Spagnolo, B., Valenti, D., Fiasconaro, A., 2004. Noise in
Ecosystems: A Short Review. Mathematical Biosciences and Engineering
1, 185-211.

\item \qquad Sprovieri, R., Di Stefano, E., Incarbona, A., Gargano, M. E., 2003. A
high-resolution record of the last deglaciation in the Sicily
Channel based on foraminifera and calcareous nannofossil
quantitative distribution. Palaeogeography, Palaeoclimatology,
Palaeoecology 202, 119-142.

\item \qquad Staliunas, K., 2001. Spatial and temporal noise spectra of spatially
extended systems with order-disorder phase transitions. Int. J.
Bifurcation and Chaos 11, 2845-2852.

\item \qquad Sugden A., Stone, R. (Editors), 2001. Special issue, Ecology Through
Time. Science 293, 623-657.

Sumedha, Martin, O.C., Peliti, L., 2007. Selection and population
size effects in evolutionary dynamics, J. Stat. Mech. P05011.

\item \qquad Tranchina, L., Bellia, S., Brai, M., Hauser, B., Rizzo, S.,
Bartolotta, A., Basile, S., 2004. Chemistry, mineralogy and
radioactivity in \emph{Posidonia oceanica} meadows from
north-western Sicily. Chemistry and Ecology 20, 203-214.

\item \qquad Tsekouras, G.A., Provata, A., Tsallis, C., 2004. Nonextensivity of
the cyclic lattice Lotka-Volterra model. Phys. Rev. E 69, 016120(7).

\item \qquad Tranchina, L., Miccich\`{e}, S., Bartolotta, A., Brai, M., Mantegna, R.
N., 2005. \emph{Posidonia oceanica} as a historical monitor device
of lead concentration in marine environment. Environmental Science
and Technology 39, 3006-3012.

\item \qquad Tuckwell, H. C., Le Corfec, E., 1998. A stochastic model for early
HIV-1 population dynamics. J. Theor. Biol. 195, 451-463.

\item \qquad Turchin, P., Oksanen, L., Ekerholm, P., Oksanen, T., Henttonen, H.,
2002. Are lemmings prey or predators? Nature 405, 562-565.

\item \qquad Valenti, D., Fiasconaro, A., Spagnolo B., 2004a. Stochastic
resonance and noise delayed extinction in a model of two competing
species. Phys. A 331, 477-486.

\item \qquad Valenti, D., Fiasconaro, A., Spagnolo, B., 2004b. Pattern formation
and spatial correlation induced by the noise in two competing
species. Acta Phys. Pol. B 35, 1481-1489.

\item \qquad Valenti, D., Schimansky-Geier, L., Sailer, X., Spagnolo, B.,
2006. Moment Equations for a Spatially Extended System of Two
Competing Species. Eur. Phys. J. B 50, 199-203.

\item \qquad Vilar, J. M. G., Sol{\'e}, R. V., 1998. Effects of Noise in Symmetric
Two-Species Competition. Phys. Rev. Lett. 80, 4099-4102.

\item \qquad Zhang, F., Li, Z., Hui, C., 2006. Spatiotemporal dynamics and
distribution patterns of cyclic competition in metapopulation. Ecol.
Model. 193, 721-735.

\item \qquad Zhonghuai, H., Lingfa, Y., Zuo, X., Houwen, X., 1998. Noise Induced
Pattern Transition and Spatiotemporal Stochastic Resonance. Phys.
Rev. Lett. 81, 2854-2857.

\item \qquad Zimmer, C., 1999. Life After Chaos. In: R. Gallagher and T.
Appenzeller (Editors), Complex Systems. Science 284, 83-86.

\end{description}

\end{document}